\documentclass[%
 reprint,
superscriptaddress,
 amsmath,amssymb,
 aps,
]{revtex4-2}

\usepackage{graphicx}
\usepackage{dcolumn}
\usepackage{bm}
\usepackage{amsmath,amsfonts,bm}
\usepackage{braket}
\usepackage{nicefrac}
\usepackage{hyperref}
\usepackage{xcolor}

\newcommand{\rhoss}{\rho^{ss}}
\newcommand{\Orhoss}{\langle O \rangle_{\rhoss}}

\newcommand{\vTheta}{\bm{\Theta}}
\newcommand{\cL}{{\cal L}}

\newcommand{\R}{\mathbb{R}}

\usepackage{xspace}
\newcommand*{\eg}{{\it e.g.}\@\xspace}

\begin{document}

\preprint{APS/123-QED}

\title{Fully differentiable optimization protocols for non-equilibrium steady states}

\author{Rodrigo A. Vargas-Hern\'andez}
\affiliation{Chemical Physics Theory Group, Department of Chemistry,University of Toronto, Toronto, Ontario, M5S 3H6, Canada.}
\affiliation{Vector Institute for Artificial Intelligence, Toronto, Canada.}
\author{Ricky T. Q. Chen}
\affiliation{Vector Institute for Artificial Intelligence, Toronto, Canada.}
\affiliation{Department of Computer Science, University of Toronto, Canada.}
\author{Kenneth A. Jung}
\affiliation{Chemical Physics Theory Group, Department of Chemistry,University of Toronto, Toronto, Ontario, M5S 3H6, Canada.}
\author{Paul Brumer}
\affiliation{Chemical Physics Theory Group, Department of Chemistry,University of Toronto, Toronto, Ontario, M5S 3H6, Canada.}

\date{\today}

\begin{abstract}
In the case of quantum systems interacting with multiple environments, the time-evolution of the reduced density matrix is described by the Liouvillian.
For a variety of physical observables, the long-time limit or steady state solution is needed for the computation of desired physical observables. 
For inverse design or optimal control of such systems, the common approaches are based on brute-force search strategies.
Here, we present a novel methodology, based on automatic differentiation, capable of differentiating the steady state solution with respect to any parameter of the Liouvillian. 
Our approach has a low memory cost, and is agnostic to the exact algorithm for computing the steady state.
We illustrate the advantage of this method by inverse designing the parameters of a quantum heat transfer device that maximizes the heat current and the rectification coefficient. Additionally, we optimize the parameters of various Lindblad operators used in the simulation of energy transfer under natural incoherent light. 
We also present a sensitivity analysis of the steady state for energy transfer under natural incoherent light as a function of the incoherent-light pumping rate. 

\end{abstract}

\maketitle


\section{\label{sec:Intro}Introduction}
Nano-scale devices are commonly described as quantum systems that interact with multiple environments or baths. 
Their performance  is usually quantified through quantum observables of the form, $\langle \hat{O}(t) \rangle = \text{Tr}[\hat{O}\rho(t)]$, where $\rho(t)$ is the reduced representation of the quantum system's state at time $t$ \cite{OQS_book}.
For systems such as quantum heat engines, batteries, and incoherently excited exciton transport systems, the observables of interest depends on the long-time limit,
\begin{eqnarray}
\langle O \rangle_{\rhoss} \equiv \lim_{t \to \infty} \langle \hat{O}(t) \rangle = \text{Tr}[ \hat{O} \rhoss],
\label{eqn:O_rhoss}
\end{eqnarray}
where  $\rhoss$ is the steady state (SS) solution which satisfies $\frac{d {\bf \rhoss}(t) }{d t} = 0$.
To solve for the steady state, the time-evolution of $\rho$ is usually described by a quantum master equation of the form,
\begin{eqnarray}
\frac{d {\bf \rho}(t) }{d t} =  F(\rho(t); \vTheta),
\label{eqn:fODE}
\end{eqnarray}
where ${F}$ is the Liouvillian, and $\vTheta \in \R^d$ represents any set of $d$-parameters used to construct the Liouvillian, \eg, bath parameters such as the decay rates for Lindblad operators, temperature(s) of the bath(s), and system-bath parameters \cite{OQS_book}. 
From the above equations, it is clear that $\rhoss$ and $\Orhoss$ both directly depend on the functional form of the Liouvillian and the value of the parameters $\vTheta$.
By knowing the gradient of $\rhoss$  with respect to any parameter of ${F}$, we could understand more in depth the effect $\vTheta$ has on $\rhoss$ and $\Orhoss$.

The simulation of nano-scale devices, through an open quantum many-body framework, has lead to the development of various algorithms, e.g., renormalization group \cite{Renorm_OQS_PRL,Renorm_OQS_PRL_2,Renorm_OQS_PRB}, meanfield methods \cite{Meanfield_OQS_PRB,Meanfield_OQS_PRX,Meanfield_OQS_arxiv}, tensor networks \cite{TN_OQS_PRA,TN_SS_PRL,TN_OQS_PRL,TN_OQS_QST,TN_ORS_NatComm}, hierarchical equations of motion\cite{Tanimura1989,Tanimura1990,Duan2017}, Heisenberg equation of motion approaches\cite{Liu2020-1,Liu2020-2}, secular and non-secular Redfield theory \cite{Redfield,Redfield_theory}, tensor transfer methods \cite{Cerrillo2014,Kananenka2016,Gelzinis2017}, and mixed quantum-classical methods \cite{Tully1998,Kapral1999,Subotnik2016}.

Over the years, many optimal control and inverse design protocols for quantum dissipative systems have been proposed  \cite{OCDS_Goerz_2014,OCDS_Rabitz_1999,OCDS_Koch_2016,OCDS_PRL_2011,OCDS_Floether_2012,AD_OQS_EPhysJ_B,AD_OQS_PRA_Jirari,AD_OQS_PRA,AD_OQS_MLST,OCDS_RL_PRA,one_photon_OQS,one_photon_OQS_2}. The goal, find the optimal set of control parameters $\mathbf{x}$ that govern the time-evolution of $\rho(t)$ by maximizing a cost function and/or a quantum observable; $\mathbf{x}^* = \textrm{arg max}_{\mathbf{x}}\; g(\mathbf{x};\rho(t))$.
Previously, the study of quantum obervables for open quantum systems ($\langle \hat{O}\rangle = \texttt{Tr}[\hat{O}\rho]$) was usually carried with either grid search or physically motivated methods \cite{OCDS_Goerz_2014,OCDS_Rabitz_1999,OCDS_Koch_2016,OCDS_PRL_2011,OCDS_Floether_2012,OCDS_PMP_PRA_2020,OCDS_PMP_2L_PRA_2007,OCDS_Ritland_2018,OCDS_PMP_PRA_2018}, mainly because numerical differentiation is prone to numerical errors and is computational inefficient for large number of parameters. Furthermore, there are only few systems that can be solved in closed form. 
Recently, with the help of machine learning tools there have been two new directions, i) control policies learned through a reinforcement learning methodology \cite{OCDS_RL_PRL,OCDS_RL_PRA,OCDS_RL_PLA,OCDS_RL_NJP}, and ii) gradient-based algorithms powered by automatic differentiation  (AD) \cite{AD_OQS_EPhysJ_B,AD_OQS_PRA_Jirari,AD_OQS_PRA,AD_OQS_MLST}.\\

The study and optimization of non-equilibrium steady state systems has only been done by brute force search or physically motivated methods, Refs. \cite{Dvira_PRE,Jung_JCP,Timur_JCP,TimurBrumer_JCP_148}.
This is due to two main difficulties, i) the need to solve for $\rhoss$ a large number of times, and ii) inefficient numerical techniques for computing $\frac{d\rhoss}{d\vTheta}$ and $\frac{d\Orhoss}{d\vTheta}$.
For the former problem, there have been recent works on how to alleviate the large cost in obtaining $\rhoss$ by parametrizing the steady state solution \cite{NN_SS_PRB_2019,NN_SS_PRL_2019,NN_SS_PRL_2_2019,AD_SS_PRE}, and determining the Liouvillian gap \cite{NN_L_gap} using various deep learning architectures. Additionally, the memory kernel has also been approximated with deep learning methodologies \cite{NN_rho_time_PRL_2019,NN_rho_time_axiv_2020,PRL_ML_nonMark_QD,JCPL_CNN_QD}.

Gradient based methods, based on $\frac{d\rhoss}{d\vTheta}$ and $\frac{d\Orhoss}{d\vTheta}$, could facilitate the navigation/search for the optimal parameters $\vTheta^*$ in the steady states where the $\Orhoss$ observable is maximum/minimum. 
While deep learning based approaches could alleviate some of the computational cost associated with obtaining $\rhoss$, none of these methodologies can improve the computation of $\frac{d\rhoss}{d\vTheta}$ since they were not designed to learn the relation between the steady state and the Liouvillian's parameters, $\rhoss=f(\vTheta)$.
The work presented here introduces a new route to efficiently compute these quantities by combining automatic differentiation \cite{AD_survey} and the implicit function theorem \citep{krantz2012implicit}.

The sections of the paper are organized as follows, Section \ref{sec:Method} presents the methodology.
Section \ref{sec:Results} contains the results and discussion for the optimization of a quantum heat transfer device and the energy transfer efficiency in a exciton model.
Lastly, the summary is in Section \ref{sec:Summary}.

\section{\label{sec:Method}Method}
In general, any physical observable that depends on the steady state is a scalar function, \eg $\langle \hat{O} \rangle_{\rhoss} = g(\rhoss, \vTheta)$ (Eq. \ref{eqn:O_rhoss}), whose gradient with respect to the parameters can be decomposed with the chain rule,

\begin{eqnarray}
\frac{\partial g(\rhoss, \vTheta)}{\partial \Theta_i} &=& 
\frac{\partial \; \langle \hat{O}(\vTheta) \rangle_{\rhoss}}{\partial \Theta_i} \nonumber \\
&=& \texttt{Tr}\left [ \left ( \frac{\partial \;\hat{O}(\vTheta)}{\partial \Theta_i} \right )\rhoss + \hat{O}(\vTheta) \left (\frac{\partial \; \rhoss }{\partial \Theta_i} \right ) \right ],
\label{eqn:grad_O_rhoss}
\end{eqnarray}
where terms of the form $\frac{\partial \;\hat{O}(\vTheta)}{\partial \vTheta}$ can be computed efficiently with automatic differentiation \cite{AD_survey}, or using closed form expressions when available.
On the other hand, the gradient of the steady state with respect to some parameters, $\frac{d\rhoss}{d\vTheta}$, is not readily available, and the standard ways of obtaining these gradients is either analytically (when possible) or via finite differences. 
However, the latter approach requires computing $\frac{d \rhoss}{d \Theta_i}$ for each single parameter separately, for a total of ${\cal O}(d)$ evaluations of the steady state, where $d$ is the number of free-parameters in the Liouvillian, $\vTheta\in\R^d$. 
This makes the finite difference approach intractable for systems that depend on a larger number of parameters. 

Computing gradients through the chain rule is the role of an automatic differentiation (AD) framework, though we note that na\"ively using AD is insufficient for computing $\frac{d\rhoss}{d\vTheta}$. 
While we can solve for $\rhoss$ by running an appropriate ODE solver for a sufficiently long period of time, differentiating through the internals of the ODE solver is prohibitive as it requires storing all intermediate quantities of the solver. 
There exists low-memory methods for computing gradients of ODE solutions but they require either the trajectory $\rho(t)$ to be stored in memory or solving $\rho(t)$ in reverse time for constant memory~\citep{chen2018neural,AD_OQS_EPhysJ_B,AD_OQS_PRA_Jirari,AD_OQS_PRA_Jirari}. 
However, the reversing approach is not applicable as the steady state, once reached, cannot be reversed.

To differentiate the steady state solution with respect to any parameter with constant memory usage, we view $\rhoss$ as the solution of a fixed point problem, 
\begin{eqnarray}
F(\rhoss; \vTheta) = 0.
\label{eqn:fixedpoint}
\end{eqnarray}
By differentiating both-sides of Eq. \ref{eqn:fixedpoint} and solving for $d\rhoss / d\vTheta$  we obtain,
\begin{equation}\label{eq:implicit_fn_thm}
    \frac{d\rhoss}{d\vTheta} = - \left( \frac{\partial F(\rhoss; \vTheta)}{\partial \rho} \right)^{-1} \left[ \frac{\partial F(\rhoss; \vTheta)}{\partial \vTheta} \right],
\end{equation}
The implicit function theorem~\citep{krantz2012implicit} (Eq. \ref{eq:implicit_fn_thm}) permits us to exactly compute $\frac{d\rhoss}{d\vTheta}$ without knowing the explicit or analytic dependence of $\rhoss$ on $\vTheta$, $\rhoss = f(\vTheta)$.
The full derivation of Eq. \ref{eq:implicit_fn_thm} is presented in Section I-A in the Supplemental Material.

For the inverse design of non-equilibrium quantum systems using gradient based methods, we require the Jacobian of $\Orhoss$ with respect to any parameter (Eq. \ref{eqn:grad_O_rhoss}).
Specifically, we only require vector-Jacobian products (VJP) of the form $\mathbf{v}^\top\frac{d\rhoss}{d\vTheta}$. In the context of computing Eq. \ref{eqn:grad_O_rhoss}, $\mathbf{v}$ is the vectorized form of $\hat{O}(\vTheta)$.
For any vector $\mathbf{v}$, the vector-Jacobian product for $\frac{d\rhoss}{d\vTheta}$ is given by the implicit function theorem as,
\begin{equation}
\label{eq:implicit_fn_thm_vjp}
    \mathbf{v}^\top\frac{d\rhoss}{d\vTheta} = - \mathbf{v}^\top \left( \frac{\partial F(\rhoss; \vTheta)}{\partial\rho} \right)^{-1} \left[ \frac{\partial F(\rhoss; \vTheta)}{\partial\vTheta} \right],
\end{equation}
where the term $\mathbf{v}^\top \left( \frac{\partial F(\rhoss; \vTheta)}{\partial \rho} \right)^{-1}$ is the solution of a linear set of equations of the form, 
\begin{equation}
\mathbf{v}^\top = \mathbf{y}^\top\left( \frac{\partial F(\rhoss; \vTheta)}{\partial \rho} \right). 
\end{equation}
The Jacobian $J_\rho := \frac{\partial F(\rhoss; \vTheta)}{\partial \rho}$ is a $m^2\times m^2$ matrix, where $m$ is the number of degrees of freedom to describe a quantum system, and it could be efficiently inverted only for small quantum systems.
Additionally, to compute $J_\rho$ using automatic differentiation, we must to evaluate the Liouvillian ${\cal O}(m^4)$ times, which could be computationally expensive for larger quantum systems.

The approach presented here avoids the computation of $J_\rho^{-1}$ by realizing that $\mathbf{v}^\top J_\rho^{-1}$ is the steady state solution of second ODE of the form,
\begin{eqnarray}
\frac{d\;\mathbf{y}}{d \;t} = \left (\frac{\partial F(\rhoss; \vTheta)}{\partial \rho} \right )\mathbf{y}^\top - \mathbf{v}^\top.
\end{eqnarray}
where the steady state solution ($\mathbf{y}^{ss}$) is,
\begin{eqnarray}
\mathbf{y}^{ss} = \mathbf{v}^\top \left(\frac{\partial F(\rhoss; \vTheta)}{\partial \rho} \right ) ^{-1}.
\end{eqnarray}
Notably, simulating this second ODE only requires vector-Jacobian products of the form $\mathbf{y}^\top (\partial F(\rhoss; \vTheta)/\partial \rho)$, that can be efficiently computed with AD. 
Thus, to differentiate the steady state solution of a QME, we essentially solve two steady state problems, both of which can be computed using any black-box steady state solver.
For all our simulations we used an adaptive step size Runge-Kutta algorithm  of order 5 \cite{Runge-Kutta} to solve for the steady state. 
Steady state solutions obtained by this approach were checked by exact long time computations.
In summary, we propose an efficient algorithm for computing $\mathbf{v}^\top\frac{d\rhoss}{d\vTheta}$ in order to allow any black-box steady state solver to be placed within an AD framework, such as JAX~\citep{jax2018github}.
For more details about AD and the implicit function theorem we refer the reader to Ref. \cite{AD_survey} and Section I-A in the Supplemental Material.

The implicit function theorem for ODE steady states was first proposed by F. J. Pineda in 1987 to generalize the back propagation algorithm for recurrent neural networks \cite{grad_RNN_PhysRevLett}. However, the methodology presented here is a generalization of that work since it does not depend on analytic forms for the Jacobian of the Liouvillian.

To summarize, we propose a method of efficiently computing exact derivatives for all parameters simultaneously with just a single evaluation of $\rhoss$, compared to finite difference approximations computed using $\mathcal{O}(d)$ evaluations.
The computation of $\frac{d\rhoss}{d\vTheta}$ is independent of how we solve for $\rhoss$.
The implementation of the proposed algorithm is available online~\footnote{\url{https://github.com/RodrigoAVargasHdz/steady_state_jax}}.
In the following sections, we carry out inverse design and sensitivity analysis by differentiating through the steady state for the optimization of quantum heat and energy transfer systems.

\begin{figure}[h!]
\centering
\includegraphics[width=0.33\textwidth]{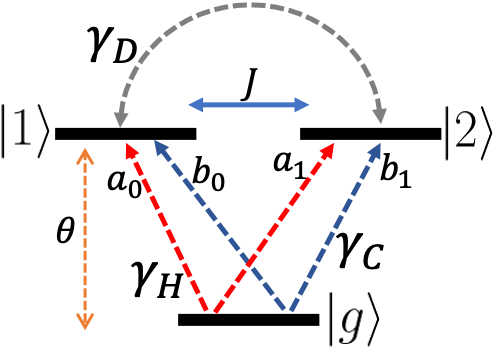}\vspace{1.0cm}
\includegraphics[width=0.33\textwidth]{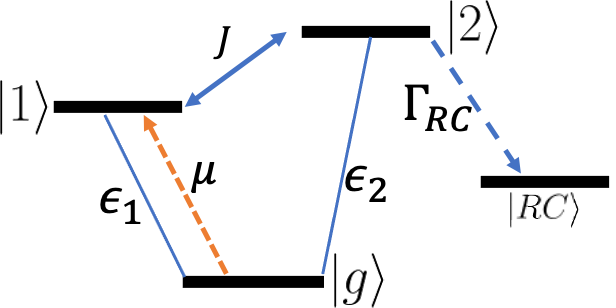}
\caption{(upper panel)  Diagram for a three level quantum heat transfer model in the local site basis. The dashed arrows represent the interactions with the hot (red) and cold (blue) bath. (lower panel) Diagram for a three level system with incoherent excitation coupled to a reaction center.}
\label{fig:Vsystems}
\end{figure}

\section{Results\label{sec:Results}}
\subsection{Redfield theory: Model systems}

Quantum heat transfer (QHT) models have been proposed as rich systems to study quantum effects in thermodynamics \cite{QHE_PRL_2019,Dvira_PRE,Goold_2016,Q_therm,QRef_PRL}.
QHT models are commonly studied within the framework of open quantum systems where various approaches of the quantum master equation  have been applied.
In the limit of weak interactions between the system and the baths, the time evolution of a QHT model can be described in a perturbative manner using Redfield theory (RT). 
RT assumes that the baths are prepared in a canonical thermal state, that the system-bath interaction can be factorized, and that the environments are Markovian. 
The Redfield master equation for multiple non-interacting baths is \cite{OQS_book,Redfield_theory},
\begin{eqnarray}
\frac{\partial \rho_{\mu,\nu}(t)}{\partial t} &=& -i\left[H_S,\rho(t)\right] + \sum_\alpha {\cal D}^\alpha\rho(t)\\
 &=& -i\omega_{\mu,\nu}\rho_{\mu,\nu}(t) + \sum_\alpha\sum_{\kappa,\lambda}R^\alpha_{\mu,\nu,\kappa,\lambda}\;\rho_{\kappa,\lambda}(t),
    \label{eqn:Redfield}
\end{eqnarray}
where ${\cal D}^\alpha$ is the dissipator for each $\alpha$ environment, and the Redfield tensor for the $\alpha$ environment is $R^\alpha$. 
In the Markovian limit and neglecting the Lamb shift, the $R^\alpha_{\mu,\nu,\kappa,\lambda}$ are,
\begin{eqnarray}
    R^\alpha_{\mu,\nu,\kappa,\lambda} &=& ^\alpha\Gamma^{+}_{\lambda,\nu,\mu,\kappa} + ^\alpha\Gamma^{-}_{\lambda,\nu,\mu,\kappa} - \delta_{\nu,\lambda}\sum_{\ell}\;^\alpha\Gamma^{+}_{\mu,\ell,\ell,\kappa} \nonumber \\
    &&- \delta_{\mu,\kappa}\sum_{\ell}\;^\alpha\Gamma^{-}_{\lambda,\ell,\ell,\nu},
    \label{eqn:Redfield_tensors}
\end{eqnarray}
where the transition rates $^\alpha\Gamma^{\pm}$ are,
\begin{eqnarray}
^\alpha\Gamma^{\pm}_{\lambda,\nu,\mu,\kappa} &=& \langle \lambda|S^\alpha |\nu \rangle \langle \mu|S^\alpha |\kappa \rangle \Upsilon^\alpha(\omega),
\label{eqn:transition_rates}
\end{eqnarray}
where $S^\alpha$ is the interaction of the system with the $\alpha-$bath and $\Upsilon^{\alpha}(\omega)$ is defined as,
\begin{eqnarray}
\Upsilon^{\alpha}(\omega) &=& \Bigg\{\begin{matrix}
\frac{1}{2}G^\alpha(|\omega|)n_\alpha(|\omega|;T_\alpha) & \omega < 0 \\ 
\frac{1}{2}G^\alpha(\omega)\left[ n_\alpha(\omega;T_\alpha) + 1\right ] & \omega > 0.
\end{matrix}
\end{eqnarray}

Each Redfield tensor depends on the system-environment interactions, the spectral density function, $G^\alpha(\omega)= \gamma_\alpha \omega e^{-\omega/\omega_c}$, and the average phonon occupation number, $n_\alpha(\omega)$. 
$[\mu,\nu,\kappa,\lambda]$ are the index of the eigenstates of $H_S$, i.e. they satisfy $H_S |\mu\ \rangle = \epsilon_{\mu} |\mu \rangle $, and $\omega_{\mu,\nu}$ is the difference between eigenvalues $\epsilon_\mu$ and $\epsilon_\nu$. 
Additionally, each environment is characterized by the temperature $T_\alpha$ and a friction coefficient $\gamma_\alpha$.
For more details about Redfield theory we refer the reader to the Supplemental Material and to Refs.
\cite{Redfield_theory,Redfield,OQS_book}.\\

Quantum heat transfer models are characterized by the change in the system's energy, $\mathrm{d} \langle H_{S} \rangle / \mathrm{d} t = \textrm{Tr}\big [ H_S \dot{\rho}(t) \big ]$, corresponding to the heat flow. This equality only holds if the system's Hamiltonian $H_{S}$ is time independent.
In the long-time limit, $\dot{\rho}(t) = 0 $, the energy exchange is comprised of the flow of heat, where the rate of heat exchange with the $\alpha$-bath is given by, 
\begin{eqnarray}
J_\alpha = \textrm{Tr}\left [H_S{\cal D}^\alpha \rhoss\right].
\label{eqn:J_q}
\end{eqnarray}
If one tries to inverse design the system, baths, or the system-baths interactions to maximize $J_\alpha$, one needs to understand the effect each parameter has in the QME, e.g., $\frac{\partial D^\alpha}{\partial \theta}$ or $\frac{\partial \rhoss}{\partial \gamma_\alpha}$.
Here, we combine automatic differentiation and Eq. (\ref{eq:implicit_fn_thm}), to inverse design the heat current using gradient based methods.
We illustrate this new methodology on a QHT model with a three-level system interacting with three baths, where the system Hamiltonian is,
\begin{eqnarray}
H_{S} &=& \epsilon_g \ket{g}\bra{g} + \theta\sum_{i=1}^2 \ket{i}\bra{i} + J\big (\ket{1}\bra{2} + \textrm{h.c.}\big).
\label{eqn:H_s}
\end{eqnarray} 
$\theta$ and $J$ describe the energy of sites $\ket{1}$ and $\ket{2}$ and the hopping between them. We set $\epsilon_g = 0$ for reference. The shorthand ``h.c.'' denotes the Hermitian conjugate of former terms in the expressions.
For the construction of the Redfield tensors we invoked the Markovian approximation, and each environment's state is described with an ohmic spectral density function and a friction coefficient $\gamma_\alpha$.
We also neglect the Lamb shift. For more details see the Supplemental Material.

The most general type of interactions with the hot (H), cold (C) and decoherence (D) baths are,
\begin{eqnarray}
S^H &=& a_0\ket{g}\bra{1} + a_1 \ket{g}\bra{2} + \textrm{h.c.} \label{eqn:_S_H}\\
S^C &=& b_0\ket{g}\bra{1} + b_1 \ket{g}\bra{2} + \textrm{h.c.}\label{eqn:_S_C}\\
S^D &=& \ket{1}\bra{2} + \textrm{ h.c.} \label{eqn:_S_D},
\end{eqnarray}
where $[a_0,a_1]$ and $[b_0,b_1]$ are the coupling strength parameters to the H and C bath, left panel Fig. \ref{fig:Vsystems}. The $D-$bath is a control mechanism to study the role of coherences between the sites.
For this system, analytic results for $J_\alpha$ can only be derived in the secular limit \cite{Dvira_PRE}. These system-bath interactions (Eqs. \ref{eqn:_S_H}--\ref{eqn:_S_D}) are needed to construct the transition rates $^\alpha\Gamma^{\pm}$ (Eq. \ref{eqn:transition_rates}).

The parameters of this QHT model can be efficiently optimized by maximizing the heat exchange with the hot bath, $J_{H}$ (Eq. (\ref{eqn:J_q})). 
For these simulations, the temperatures are held fixed for all three baths, $T_{H} = 0.15$, $T_{C} = 0.1$, and $T_{D} = 0.12$.
The final space of parameters is $\vTheta = [\theta,J,a_0,a_1,b_0,b_1,\gamma_{H},\gamma_{C},\gamma_{D}]$. 

For each optimization procedure, all initial parameters were randomly sampled.
Values of $a_i$ and $b_i$ were constrained to $[0,1]$ to avoid physically incorrect models, and the values of $\theta$ and $J$ were constrained to the positive domain by casting them as the exponential function of unconstrained variables.
To maximize $J_{H}$, we used the Adam \cite{Adam} optimization algorithm with a learning rate of 0.02.

\begin{figure}
\centering
\includegraphics[width=0.45\textwidth]{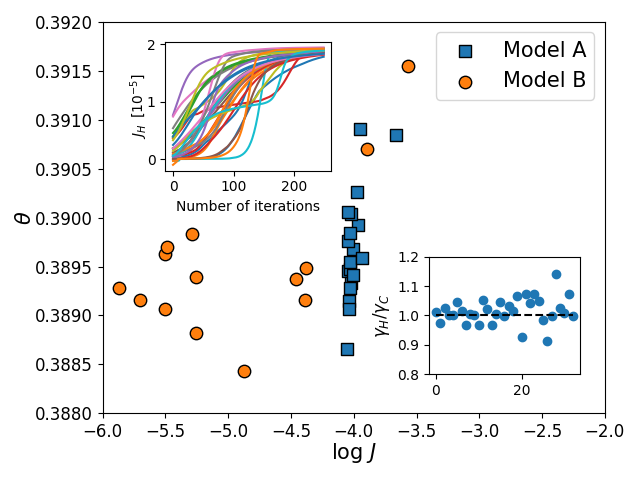}
\caption{The symbols in the main and lower inset panels represent the optimal set of parameter that maximize the heat exchange for the hot bath ($J_{H}$), Eq. \ref{eqn:J_q}, for a quantum heat transfer system, (Fig. \ref{fig:Vsystems}).
We categorize the optimal set of parameter into two different cases, i) model A for $a_0 = b_1 \approx 1$ or $a_1 = b_0 \approx 1$, and ii) model B for $a_i = b_i \approx 1$. Model A the hot (H) and cold (C) baths interact with different sites, and for model B the H and C baths interact with the same site. 
All parameters,  $\vTheta = [\theta,J,a_0,a_1,b_0,b_1,\gamma_{H},\gamma_{C},\gamma_{D}]$, were optimized with the Adam algorithm.
The upper inset figure depicts the value of the heat exchange during the optimization procedure, where the initial values of $\vTheta$ were sampled randomly. We only considered 100 random initializations of $\vTheta$. 
 }
\label{fig:QHE_1}
\end{figure}

In Ref. \cite{Dvira_PRE}, this system was studied using fixed parameters $a_0 = b_1 = 1$ and $a_1 = b_0 = 0$. Through optimization, we found that the majority of the optimized systems also recovered these parameter values (Model A). 
However, the remainder of the optimized results indicate that $J_{H}$ is maximum when both baths, hot and cold, interact with only one site, e.g., $a_0 = b_0 \approx 1$ while $a_1 = b_1 \approx 0$ or vice-versa (Model B).
Fig. \ref{fig:QHE_1} contains a \emph{collection} of optimizations, with the optimized values of $\theta$ and $J$ indicated and those of the remaining parameters not explicitly shown. 
In addition, the upper inset shows that all of these optimizations converge to essentially the same value of $J_{H}$.
It is worth noticing that independent of the initial parameters, $J_{H} > 1.85\times10^{-5}$ after 200 iterations; Fig. \ref{fig:QHE_1} upper inset panel. 

Our QHT model considers any linear combination of interactions between both hot and cold baths with any site. 
It is not a surprise that $J_{H}$ is maximized when $a_i \approx 0$ or $b_i \approx 1$, since a stronger system-bath interaction increases the heat transfer. 
However, by being able to independently optimize $a_i$ and $b_i$, we found that the magnitude of the heat exchange for Model B is the same as for Model A, making it an interesting future route for further examination since it has never been studied. 

We stressed that the parameters for both Models A and B where found by maximizing $J_{H}$ with Adam \cite{Adam}, a first order gradient descent method. 
The optimal value of the $J$ parameter for Model A is $J \approx 0.018$, and for Model B $0.0025 < J < 0.03$. 
For both systems, Model A and B, the optimal value of the energy of sites $\ket{1}$ and $\ket{2}$ is $\theta \approx 0.3895$.
The optimized parameters are reported in Table I in the Supplemental Material.

$\gamma_H$ and $\gamma_C$ are the friction parameters that describe the strength of the coupling to each bath. For each individual optimization we found that at the end of the search procedure, $\gamma_H/\gamma_C\approx 1$. This is an interesting property that can depend on the fixed values of the temperatures. 
We also found that the value of $\partial J_{H} / \partial \gamma_D$ is zero in the regions  where $J_{H}$ is maximum. 
Given the weak-interacting assumption inherent within Redfield theory, the maximum value allowed for $\gamma_i$ is 0.0025 and the initial values for these parameters were only sampled from $[10^{-5},10^{-4}]$. The gradient of $\partial J_{H} / \partial \gamma_H$ and $\partial J_{H} / \partial \gamma_C$ show that their values must increase in order to maximize the heat exchange.

For this case, each iteration requires only two steady state evaluations with our approach, while a finite difference approach would have needed $2 \times |\vTheta|$ = 18 evaluations of the steady state. Additionally, each gradient-based optimization took less than 150 iterations to find optimal parameters with high precision, Fig. \ref{fig:QHE_1}. A standard grid-search approach of 10 points for each parameter would have required over $10^9$ steady state evaluations.

We also considered a quantum heat transfer model with non-degenerate states,
\begin{eqnarray}
H_{S} &=& \epsilon_g \ket{g}\bra{g} + \sum_{i=1}^2 \theta_i\ket{i}\bra{i} + J\big(\ket{1}\bra{2} + \textrm{h.c.}\big),
\end{eqnarray}
using the same procedure and the same system-bath interactions, Eqs. \ref{eqn:_S_C}--\ref{eqn:_S_D}. We found that degenerate systems, $\theta_1 /\theta_2 \approx 1$ are the systems with the highest $J_H$. 
The optimal values of using separate parameters for each site ($\theta_1$ and $\theta_2$) were similar to the optimized value of using a single on-site energy parameter ($\theta$) to describe both sites in Eq. \ref{eqn:H_s}. The optimal value found was $\theta_i = \theta  \approx 0.389$ (Fig. \ref{fig:QHE_1}). 
For these degenerate Hamiltonians it was found that the hot and the cold bath interacting with different sites (Model A) was ideal. It is well known that degeneracy between sites leads to increased coherences in transport systems which in turn increases the currents. This demonstrates that even with limited input the method outlined in Sec. \ref{sec:Method}  can lead to correct physical models.
All parameters are reported in Table II in the Supplemental Material.\\

\begin{figure}
\centering
\includegraphics[width=0.45\textwidth]{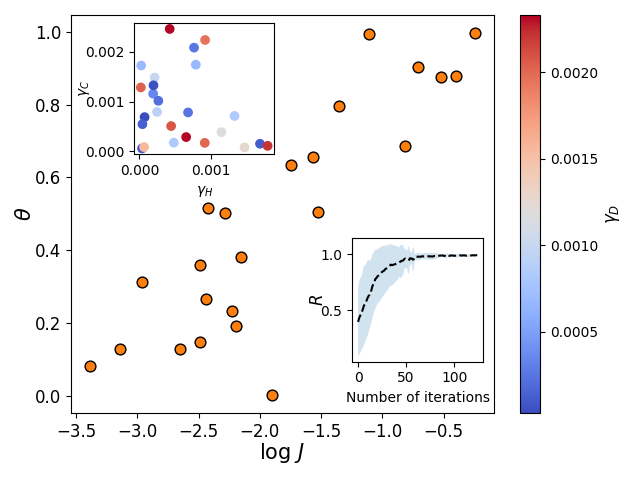}
\caption{The symbols in the main and upper inset panels represent the optimal set of parameter that maximize the rectification coefficient ($R$), Eq. \ref{eqn:R}, for a quantum heat transfer system, (Fig. \ref{fig:Vsystems}).
All parameters,  $\vTheta = [\theta,J,\gamma_H,\gamma_C,\gamma_D]$, were optimized with the Adam algorithm. The hot and cold bath only interact to different sites.
The lower inset figure depicts the averaged value of the rectification coefficient during the optimization procedure for 25 random initialization of $\vTheta$. 
\label{fig:QHE_R}}
\end{figure}

Another common observable to study quantum heat transfer models is the rectification coefficient, $R$, which is the net heat current when the temperature difference, $T_{H} - T_{C}$, in the H and C reservoirs is reversed \cite{Motz_2018},
\begin{eqnarray}
R = \frac{|J_{H}| - |J'_{H}|}{|J_{H}| + |J'_{H}|},
\label{eqn:R}
\end{eqnarray}
where $J_{H}$ is the heat current when the hot bath interacts with $\ket{1}$ and cold bath with  $\ket{2}$, and $J'_{H}$ is computed by swapping the temperatures of the H and C bath. For all simulations, we fixed the values $a_i$ and $b_i$ to $a_0=b_1=1.$ and $a_1=b_0=0$, and we again held fixed $T_{H} = 0.15$, $T_{C} = 0.1$, and $T_{D} = 0.12$.
The same methodology used to optimize a quantum heat transfer system where the surrogate observable was $J_{H}$ can be used to tune the free parameters for $R$. 
We use the same gradient-based algorithm, Adam, to maximize $R$. For these simulations, we only considered as free parameters, $\theta,J$ and the friction coefficients for all three baths, $[\gamma_{H},\gamma_{C},\gamma_{D}]$. 
Results of the optimal parameters, as in Fig. \ref{fig:QHE_1} for a set of obtained cases, are displayed in Fig. \ref{fig:QHE_R}. 
As stressed through this paper, inverse designing quantum heat transfer model with modern gradient methods like Adam, requires very few number of iterations---roughly $100$ or less for this case---to find optimal parameter values.

For each individual optimization, we first sampled the parameters $\theta,J,\gamma_{H},\gamma_{C},\gamma_{D}$ and used Adam to find the maximizer of $R$. 
All optimizations were stopped once $R>95\%$, which on averaged took approximately 100 iterations; lower inset in Fig. \ref{fig:QHE_R}.
Our results illustrate that there is a linear trend between $\theta$ and $J$.
However, the optimal range of $\theta$ when $R$ is maximum is wider than for $J_{H}$, indicating that both quantum observables do not share the same set of optimal physical parameters.
\\

\subsection{\label{sec:KennyModel}Energy transfer for the V-system}
As a second system, we consider a simplified model of energy transfer shown in the right panel in Fig. \ref{fig:Vsystems}. When the radiation incident on the donor state $\ket{1}$ is taken to originate from an incoherent source, such as the sun, the system is a useful minimal model of biological energy transfer. 
The steady state efficiency, $\eta_{loc}$, is quantified by;
\begin{eqnarray}
\eta_{loc} &=& \frac{\Gamma_{RC}}{r}\rhoss_{2}. 
\label{eqn:KennyModel_eff}
\end{eqnarray}
Here $\rhoss_{2}$ is the probability of being in the site neighboring the reaction center, ($\ket{2}$), Fig. \ref{fig:Vsystems}, $\Gamma_{RC}$ is the rate of energy transfer from the acceptor state $\ket{2}$ to the reaction center $\ket{RC}$, and $r$ is the  incoherent-light pumping rate. 
The time evolution of this system is modeled by,
\begin{eqnarray}
\frac{\partial \rho_{S}}{\partial t} = \cL_{0}[\rho] + \cL_{rad}[\rho] + \cL_{deph}[\rho] + \cL_{rec}[\rho] + \cL_{RC}[\rho],
\label{eqn:qme_KM}
\end{eqnarray}
where the first term is the unitary evolution of the system, $\cL_{0}[\rho] = -i \left [ H_{S}, \rho \right ]$, and the rest of the terms, ${\cal L}_i$, describe the radiation (rad), the trapping of the excitons at the reaction center (RC), environmental dephasing (deph), and the recombination of the excitons (rec). See the Supplemental Material and Ref. \cite{Jung_JCP} for more information. 

\begin{figure}
\includegraphics[width=0.45\textwidth]{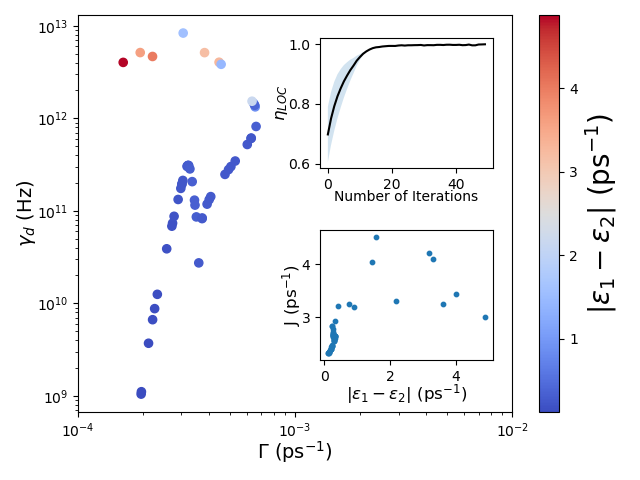}
\caption{Optimal parameters found by maximizing $\eta_{loc}$ using Adam. We considered different random initializations of $\Gamma$, $\gamma_d$, $J$ and $|\epsilon_1 - \epsilon_2|$. Each optimize model has a $\eta_{loc}\approx 1.$. 
The optimal values of $\Gamma$, $\gamma_d$, $J$ and $|\epsilon_1 - \epsilon_2|$ are presented in the main and inset lower panels. 
For these simulations, we fixed the values of $\Gamma_{RC} = 0.5$ ps$^{-1}$, $|\epsilon_1 - \epsilon_2| = 1.3$ ps$^{-1}$, and $r = 6.34\times10^{-10}$ ps$^{-1}$ \cite{Timur_JCP}.
The upper inset panel presents $\eta_{loc}$ averaged throughout the optimization procedure for different random initialization. 
\label{fig:Kmodel}}
\end{figure}

\begin{figure}
\centering\includegraphics[width=0.45\textwidth]{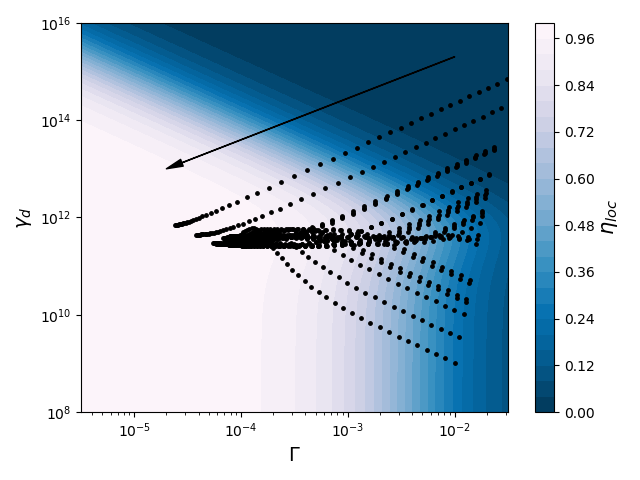}
\centering\includegraphics[width=0.45\textwidth]{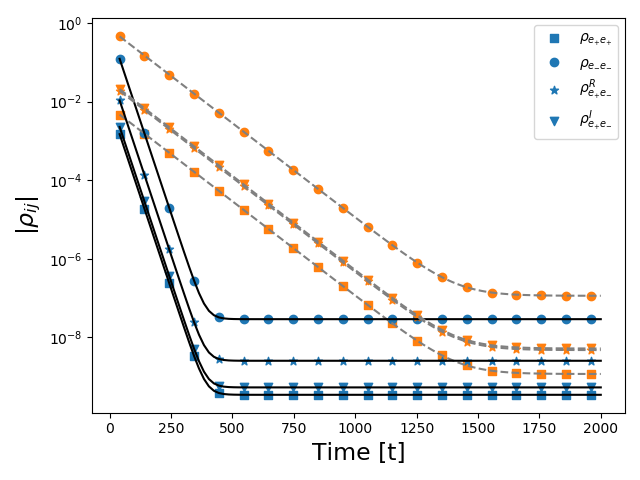}
\caption{(upper panel) Optimization steps for $\eta_{loc}$ for different random initializations, starting in the right with $\eta_{loc} \sim 0$. The markers represent the optimization trajectories that the Adam algorithm follows. (lower panel) The time evolution of $\rho$ until a steady state is reached. We considered two cases, a random initial parameters (dashed curves) where $\eta_{loc}\approx 0.0$, and for optimized parameters (solid curves) where $\eta_{loc}\approx 1$.
The random initial parameters considered were, $\gamma_d = 2.88\times 10^{12}$ Hz and $\Gamma = 0.0194$ ps$^{-1}$, and the optimized ones are $\gamma_d = 3.53\times 10^{11}$ Hz  and $\Gamma = 7.2\times 10^{-5}$ ps$^{-1}$.
For these simulations, we fixed the values of $\Gamma_{RC} = 0.5$ ps$^{-1}$ and $r = 6.34\times10^{-10}$ ps$^{-1}$ \cite{Timur_JCP}.
\label{fig:Kmodel_Adam}}
\end{figure}

We optimize  $\vTheta = [\Gamma,\gamma_d,|\epsilon_1 - \epsilon_2|,J]$ by maximizing $\eta_{loc}$ using the Adam algorithm, Figs. \ref{fig:Kmodel}--\ref{fig:Kmodel_Adam}.
$|\epsilon_1 - \epsilon_2|$ is the energy difference between site $\ket{1}$ and $\ket{2}$, $J$ is the hopping amplitude, both being system parameters. $\Gamma$  corresponds  to  the  recombination  rate, and $\gamma_d$ is the phonon bath dephasing rate. 
For each optimization, we randomly sampled different values for the parameters. 
We fixed the values of $\Gamma_{RC}$ to $\Gamma_{RC} = 0.5$ ps$^{-1}$ and the incoherent-light pumping rate $r = 6.34\times10^{-10}$ ps$^{-1}$; parameters taken from Ref. \cite{TimurBrumer_JCP_148}. 
Optimal set of parameters are presented in Fig. \ref{fig:Kmodel} and Table III in the supplemental material.

From Fig. \ref{fig:Kmodel}, we can observe that when the value of $J$ is small, there is a linear correlation with the difference between the energy sites,  $|\epsilon_1 - \epsilon_2|$. 
For $\gamma_d$ the optimal possible values span a wider range, from $10^{9}$ to $10^{13}$ Hz; however, for larger values of $\gamma_d$ the $|\epsilon_1 - \epsilon_2|$ optimal parameters must be greater as well. 
The optimal range for $\Gamma$ was less wider than the rest, and interestingly, this range is aligned with the values used in physical simulations of Ref. \cite{Timur_JCP}.
For these simulations, the initial parameters were sampled from a region where $\eta_{loc}$ is not optimal, however, our approach managed to optimize the parameters regardless of their initial values.
Throughout our optimizations, we notice that it only took approximately 20 iterations to reach $\eta_{loc}>99\%$ (Fig. \ref{fig:Kmodel} upper inset and left panel in Fig. \ref{fig:Kmodel_Adam}). 
As we can observe from the left panel of Fig. \ref{fig:Kmodel}, for some values of $|\epsilon_1 - \epsilon_2|$ and $J$ the plateau where $\eta_{loc}$ is maximum is when $\Gamma$ and $\gamma_d$ have small values. 
We report all optimal parameters in Table III in the Supplemental Material. 

In Fig. \ref{fig:Kmodel_Adam} we display the optimization trajectories, using Adam, where all random initial parameters had $\eta_{loc}<20\%$ and the end result where $\eta_{loc}>99\%$. 
The time evolution of $\rho(t)$ for a pair of random and optimal set of parameters is display in the right panel of Fig. \ref{fig:Kmodel_Adam}.
The random initial parameters values are $\gamma_d = 2.88\times 10^{12}$ Hz and $\Gamma = 0.0194$ ps$^{-1}$, and the optimized ones are $\gamma_d = 3.53\times 10^{11}$ Hz  and $\Gamma = 7.2\times 10^{-5}$ ps$^{-1}$. For these simulations we fixed the value for the other parameters to $\Gamma_{RC} = 0.5$ ps$^{-1}$, $|\epsilon_1 - \epsilon_2| = 1.3$ ps$^{-1}$, and $r = 6.34\times10^{-10}$ ps$^{-1}$ \cite{Timur_JCP}.
As it can be observed, there is a significant difference in $\rhoss$ for the random initial parameters and the optimized ones.

\begin{figure}
\centering
\includegraphics[width=0.45\textwidth]{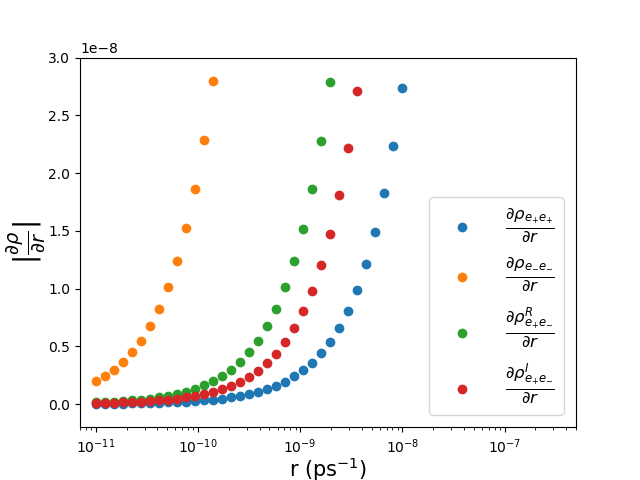}
\caption{Sensitivity analysis of $\rhoss$ with respect to the pumping rate, $r$. We compute $\partial \rhoss_{i,j} / \partial r$ using Eq. \ref{eqn:fixedpoint}. For this calculations we used $\Gamma_{RC} = 0.5$ ps$^{-1}$, $\gamma_d = 1$ Hz, $J = 0.12$ ps$^{-1}$, and $|\epsilon_1 - \epsilon_2| = 1.3$ ps$^{-1}$.}
\label{fig:Kmodel_dross_dr}
\end{figure}

As we pointed above, our algorithm allows us to compute the vector-Jacobian product of the steady state with respect to any parameter of the Liouvillian.
So far, the main application has been the inverse design of open quantum systems by maximizing/minimizing quantum observables using gradient based methods.
However, the Jacobian can also be used to understand the effect a set of parameters have on a quantity of interest, sensitivity analysis. 
For non-equilibrium steady state systems and before this work, the Jacobian of the steady state was only computed when analytic solutions were available, or by finite differences. 
Here, as the last example, we study the effects of the attenuation of the incident radiation which is important since photon absorbing centers are found in a variety of environments \cite{Axelrod2018,Axelrod2019,Chuang2020}.
By computing $\partial \rhoss / \partial r$ we found that $\rhoss_{e_{-},e_{-}}$ is the most sensitive to $r$, i.e. $\partial \rhoss_{e_{-},e_{-}}/\partial r > \partial \rhoss_{e_{+},e_{+}}/\partial r$, Fig. \ref{fig:Kmodel_dross_dr}.  Here, $e_\pm$ are the eigenstates of the system Hamiltonian, $H_{S} \ket{e_\pm} = \epsilon_{\pm}\ket{e_\pm}$, and $\rhoss_{e_{\pm},e_{\mp}}$ are the matrix elements of the steady state density matrix in the eigenbasis.
Additionally, from Fig. \ref{fig:Kmodel_dross_dr} we can also observe that the imaginary part of the coherence, $\Im(\rho_{e_{+},e_{-}})$, which relates to the exciton flux \cite{Jung_JCP}, is less sensitive until $r \approx 0.5 10^{-9}$ ps$^{-1}$, indicating that the pumping rate will not increase the flux below $r \sim 0.5 10^{-9}$ ps$^{-1}$.\\

\section{\label{sec:Summary} Summary}
Inverse design and optimal control protocols for open quantum systems that are quantified through observables that depend on the steady state, $\Orhoss$, are commonly done with brute-force search or inspired methods. 
On the other hand, gradient-based algorithms have proven to be efficient tools to minimize/maximize functions. 
For non-equilibrium steady state systems, the technical limitation was the inability to efficiently compute the Jacobian of the steady state with respect to any parameter of the Liouvillian; $\frac{d\rhoss}{d\vTheta}$.
We circumvent this by combining automatic differentiation and the implicit function theorem.
Furthermore, we believe that the present work is the first example of the application of gradient-based methods to efficiently inverse design non-equilibrium steady state systems. 
All systems were optimized with Adam, a first order gradient algorithm; however, the procedure proposed here, combined with automatic differentiation, can be also applied to efficiently compute the Hessian matrix, used in second order gradient  optimization algorithms.

The optimal design of non-equilibrium systems is still driven by physical intuition. 
However, with the possibility to compute $\frac{d\rhoss}{d\vTheta}$, we could engineer more robust systems, baths, and system-bath interactions.  
Furthermore, this methodology could also be used to study the sensitivity of $\rhoss$ with respect to any parameter in the Liouvillian, and, for example, could lead to more insight in how light affects biological processes.

Any inverse design protocol must create physically valid parameters. While this could be taken into account by adding some constraints to the main physical observable of interest, here we decided to take a different route by leveraging the flexibility of automatic differentiation. For example, to constrain the system-bath parameters to $[0,1]$ we used the soft-max function \cite{softmax_Bridle}, and for $\gamma_i$'s the range was constrained to $[0,0.0025]$, the soft-max function times the maximum value allowed. 
By constraining the range of $a_i$, $b_i$, and $\gamma_i$, the optimization of the system remains in the  weak-interacting limit where the Redfield theory is valid.
Similar algebraic transformations could be applied to constrain the value of other parameters to ensure a valid physical range for experimental setups.

In the cases introduced, optimization of a single target quantity (e.g. the heat transfer, or energy efficiency) was carried out in models parametrized by several quantities. As a result, numerous optimized models were obtained, each with similar values of the optimized target. That is, interestingly, the parameter surface has multiple maxima of similar depth. In cases where one is attempting to achieve optimization of multiple quantities (e.g. heat transfer \emph{and} verification) the proposed methodology could be integrated into gradient-based algorithms for multi-objective optimization to construct the Pareto front \cite{MGDA}.

In all the simulations presented here, an ODE solver was used to obtain $\rhoss$; however, Eq. \ref{eq:implicit_fn_thm} is agnostic to the exact method used to obtain $\rhoss$. This makes the present methodology particularly valuable. Additionally, this method could be used to study any other system whose observables depend on a long-time solution for equations of the form of Eq. \ref{eqn:fODE}. Finally, this methodology opens the possibility of studying more complex quantum heat transfer devices or natural-light induced processes.\\

 A tutorial of the method presented here is publicly available in the repository: \url{https://github.com/RodrigoAVargasHdz/steady_state_jax}

We acknowledge fruitful discussions with Professor Dvira Segal. 
This work was funded by components of two grants from the  US Air Force Office of Scientific Research, FA9550-19-1-0267 and FA9550-20-1-0354.

\bibliography{references}
\end{document}



\title{Supplemental material for "Fully differentiable optimization protocols for non-equilibrium steady states"}

\author{Rodrigo A. Vargas-Hern\'andez}
\affiliation{Chemical Physics Theory Group, Department of Chemistry,University of Toronto, Toronto, Ontario, M5S 3H6, Canada.}
\affiliation{Vector Institute for Artificial Intelligence, Toronto, Canada.}
\author{Ricky T. Q. Chen}
\affiliation{Vector Institute for Artificial Intelligence, Toronto, Canada.}
\affiliation{Department of Computer Science, University of Toronto, Canada.}
\author{Kenneth A. Jung}
\affiliation{Chemical Physics Theory Group, Department of Chemistry,University of Toronto, Toronto, Ontario, M5S 3H6, Canada.}
\author{Paul Brumer}
\affiliation{Chemical Physics Theory Group, Department of Chemistry,University of Toronto, Toronto, Ontario, M5S 3H6, Canada.}

\date{\today}

\begin{abstract}
The  purpose  of  this  supplemental  material  is  to  provide  details  of  the  numerical  calculations  we  present  in  this work. 
Section \ref{sec:AD} presents the core of the algorithm, and
sections \ref{sec:Lindblad} and \ref{sec:Redfield} describe the details of the quantum systems used in the main work. 
In section \ref{sec:Results} we present numerical solutions of the optimal parameters presented in the main text.
\end{abstract}

\keywords{Suggested keywords}
\maketitle

\section{\label{sec:AD}Automatic differentiation}
Automatic differentiation (AD) is a framework designed to compute exact vector-Jacobian ($\mathbf{v}^\top J$) and Jacobian-vector ($J\mathbf{v}$) products \cite{AD_survey}, where $J$ is any Jacobian and $\mathbf{v}$ is a vector.
The former is often referred to as reverse-mode automatic differentiation, and the latter as forward-mode.
For a function $\mathbf{f}: \R^n \to \R^m$ the full Jacobian evaluated at an input value $\mathbf{x}$ is a matrix of $m$ by $n$:
\begin{eqnarray}
J(\mathbf{x}) &=&  \begin{bmatrix}
\frac{\partial \mathbf{f}}{\partial x_1}, & \cdots, & \frac{\partial \mathbf{f}}{\partial x_n}\\ 
\end{bmatrix} 
=\begin{bmatrix}
\frac{\partial f_1}{\partial x_1}, & \cdots,  & \frac{\partial f_1}{\partial x_n} \\ 
 \vdots, & \frac{\partial f_i}{\partial x_j}, &  \vdots\\ 
\frac{\partial f_m}{\partial x_1},  & \cdots, & \frac{\partial f_m}{\partial x_n}
\end{bmatrix},
\end{eqnarray}
where the row vector is a shorthand notation for the matrix.

For scalar functions, $\mathbf{f}: \R^n \to \R^1$ it is clear that with a single vector-Jacobian product it is possible to construct the entire Jacobian. For this case, $\mathbf{v}$ is the scalar 1. In our setting, the optimization objective is a quantum observable, $\langle \hat{O} \rangle = \text{Tr\;} [\hat{O}  \rho]$, which is a scalar function. Thus, with reverse-mode AD, it is possible to construct the entire Jacobian with the same asymptotic cost as a single computation of $\langle \hat{O} \rangle$.
\begin{eqnarray}
\frac{d \langle \hat{O} \rangle}{d \boldsymbol{\Theta}}  = \begin{bmatrix}
\frac{\partial \langle \hat{O} \rangle}{\partial \Theta_1},  & \cdots, & \frac{\partial \langle \hat{O} \rangle}{\partial \Theta_n} 
\end{bmatrix}
\label{eqn:AD_back_QObs}
\end{eqnarray}

In this work we take advantage of efficient gradient computation to illustrate that complex quantum observables could be numerically optimized by combining AD and  gradient-based optimization methods such as Adam \cite{Adam}. 
Since quantum observables are often defined as a function of the steady state, we require efficient evaluation of $\frac{d \rho^{ss}}{d \boldsymbol{\Theta}}$, where $\rho^{ss}$ is the steady state solution, $\frac{d {\bf \rho}(t) }{d t} = 0$, of a first-order differential equation, 
\begin{eqnarray}
\frac{d {\bf \rho}(t) }{d t} = F(\rho(t); \boldsymbol{\Theta}).
\label{eqn:fODE}
\end{eqnarray}
$\vTheta$ is any set of parameters of $F(\cdot)$, and $\rho$ is the reduced density matrix. 
Here we studied the evolution of quantum systems that are affected by multiple external environments, and $\vTheta$ can contain controllable parameters of the environment. 

\subsection{\label{sec:Fix_Point}Fixed-point problem: computing the Jacobian for the steady state solution }
In the steady state limit, $\frac{d {\bf \rho}(t) }{d t} = 0$, we can recognize that, 
\begin{eqnarray}
F(\rho^{ss}; \boldsymbol{\Theta}) = 0.
\end{eqnarray}
Conditioned on a given set of parameters $\boldsymbol{\Theta}_0$, we denote $\rho^{ss}$ as the solution that satisfies the above equation.
By differentiating with respect to $\boldsymbol{\Theta}_0$ and evaluating at the point $(\rho^{ss},\boldsymbol{\Theta}_0)$ we get,
\begin{eqnarray}
\frac{\partial F(\rho^{ss},\boldsymbol{\Theta}_0)}{\partial \boldsymbol{\Theta}} + \frac{\partial  F(\rho^{ss},\boldsymbol{\Theta}_0)}{\partial \rho} \frac{d \rho^{ss}}{d \boldsymbol{\Theta}_0}  = 0.
\label{eqn:deriv_fixedpoint}
\end{eqnarray}
We can re-arrange this equation and obtain an expression for the term of interest:
\begin{eqnarray}
\frac{d \rho^{ss}}{d \boldsymbol{\Theta}_0} = - \Bigg[\frac{\partial  F(\rho^{ss},\boldsymbol{\Theta}_0)}{\partial \rho}  \Bigg]^{-1}\frac{\partial F(\rho^{ss},\boldsymbol{\Theta}_0)}{\partial \boldsymbol{\Theta}}.
\end{eqnarray}
Any of the terms in Eq. \ref{eqn:deriv_fixedpoint} can be computed exactly with AD. 
In other words, given an $\boldsymbol{\Theta}_0$ no matter how we solve for $\rho^{ss}$, we can still compute the Jacobian using just derivative information at the solution point $\vTheta_0$.
Also, the above procedure is agnostic to how the steady state is solved, and avoids differentiating through the ODE solver. The memory cost is thus constant and does not scale with the complexity of the problem.

\begin{figure}
\centering
\includegraphics[width=0.75\textwidth]{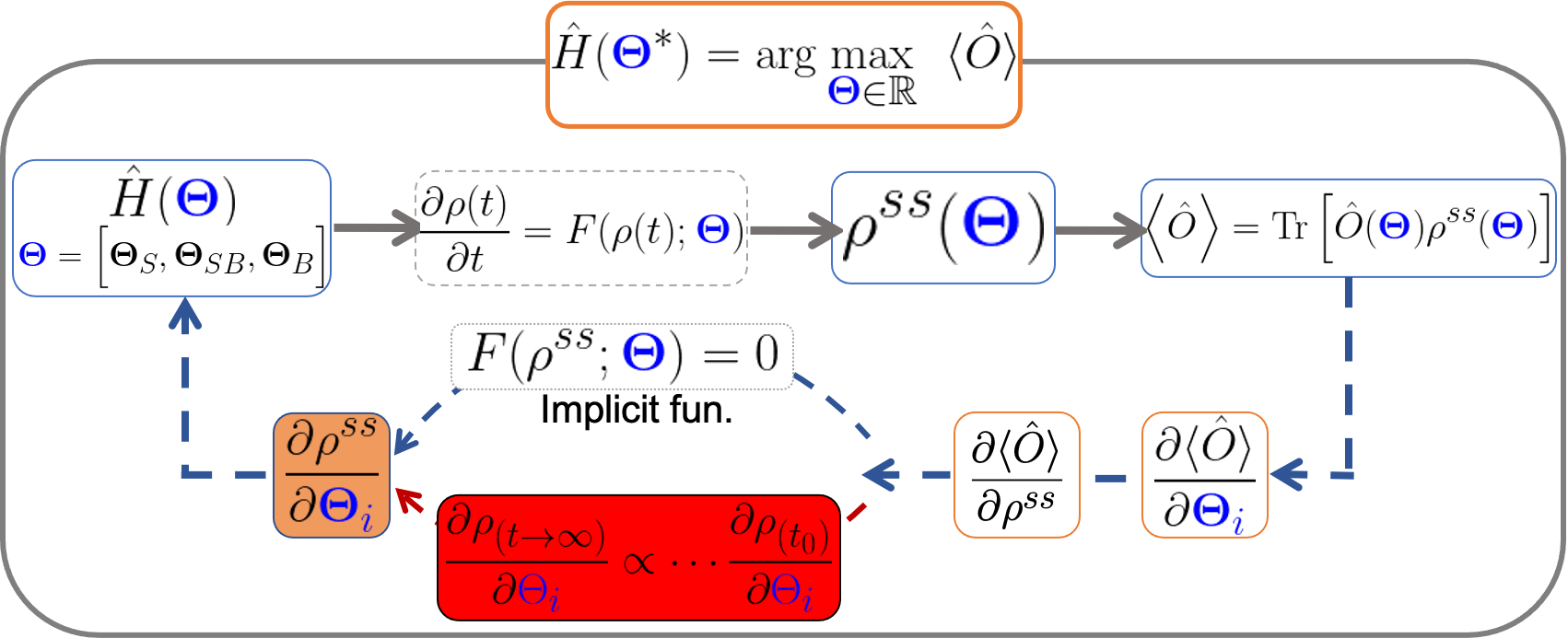}
\caption{Diagram of the methodology proposed here. $\vTheta$ represents the entire set of parameters of $F(\cdot)$.}
\label{fig:AD_rhoss_diagram}
\end{figure}

All the simulations in this work were coded in the suite JAX \cite{jax2018github}.
Figure \ref{fig:AD_rhoss_diagram} is the computational diagram of the methodology, presented in the main work, to compute $\frac{d\rhoss}{d\vTheta}$, where $\vTheta$ is the collection of parameters of $F(\cdot)$.

\section{\label{sec:level1}Model descriptions}


The time-evolution of systems of interest in this work are described within an open quantum system formalism.
We describe the steady state of the  quantum heat transfer using Redfield theory \cite{Redfield_theory,Redfield}, and for the energy transfer model for the donor-acceptor excitonic model we applied Lindblad \cite{OQS_book,Lindblad_QME_JCP,Lindblad1976}.
Both systems and theories are briefly presented in the following sections.
We optimize the parameters of both models using a gradient descent method (Adam).

\subsection{\label{sec:Redfield}Heat transfer model}

The system Hamiltonian for a quantum heat transfer model is,
\begin{eqnarray}
H_{S} &=& \epsilon_g \ket{g}\bra{g} + \theta\sum_{i=1}^2 \ket{i}\bra{i} + J\big (\ket{1}\bra{2} + \textrm{h.c.}\big),
\end{eqnarray} 
where $\theta$ and $J$ describe the energy of sites $\ket{1}$ and $\ket{2}$ and the hopping between them. We set $\epsilon_g = 0$ for reference. This system will be interacting with two baths and is
treated using Redfield theory (RT), which is a second-order time-local master equation \cite{OQS_book,Redfield_theory}. 
For multiple non-interacting baths with factorized initial conditions the Redfield quantum master equation for states $\ket{g}$ and $\ket{i}$,
\begin{eqnarray}
\frac{\partial \rho_{\mu,\nu}(t)}{\partial t} &=& -i\left[H_S,\rho(t)\right] + \sum_\alpha {\cal D}^\alpha\rho(t)\\
 &=& -i\omega_{\mu,\nu}\rho_{\mu,\nu}(t) + \sum_\alpha\sum_{\kappa,\lambda}R^\alpha_{\mu,\nu,\kappa,\lambda}\;\rho_{\kappa,\lambda}(t),
    \label{eqn:Redfield}
\end{eqnarray}
where ${\cal D}^\alpha$ is the dissipator for each $\alpha$ environment, and the Redfield tensors for the $\alpha$ environment $R^\alpha$, are,
\begin{eqnarray}
    R^\alpha_{\mu,\nu,\kappa,\lambda} &=& ^\alpha\Gamma^{+}_{\lambda,\nu,\mu,\kappa} + ^\alpha\Gamma^{-}_{\lambda,\nu,\mu,\kappa} - \delta_{\nu,\lambda}\sum_{\ell}\;^\alpha\Gamma^{+}_{\mu,\ell,\ell,\kappa} \nonumber - \delta_{\mu,\kappa}\sum_{\ell}\;^\alpha\Gamma^{-}_{\lambda,\ell,\ell,\nu},
    \label{eqn:Redfield_tensors}
\end{eqnarray}
where $^\alpha\Gamma^{\pm}$ are the transition rates,
\begin{eqnarray}
^\alpha\Gamma^{+}_{\lambda,\nu,\mu,\kappa} &=& \langle \lambda|S^\alpha |\nu \rangle \langle \mu|S^\alpha |\kappa \rangle \int_{0}^{\infty} \mathrm{d} \tau\; \Omega_\alpha(\tau)e^{-i\omega_{\mu,\kappa}\tau}, \\
^\alpha\Gamma^{-}_{\lambda,\nu,\mu,\kappa} &=& \langle \lambda|S^\alpha |\nu \rangle \langle \mu|S^\alpha |\kappa \rangle \int_{0}^{\infty}  \mathrm{d} \tau  \;\Omega^*_ \alpha(\tau)e^{-i\omega_{\lambda,\nu}\tau}.
\label{eqn:gamma_tensors}
\end{eqnarray}
$\Omega_\alpha(\tau)$ is the correlation function for the $\alpha$-bath, and $\langle \lambda|S^\alpha |\nu \rangle$ are the matrix elements of the system-bath interaction ($S$),
\begin{eqnarray}
\Omega_\alpha(\tau) = \int_{0}^{\infty} \mathrm{d} \omega \; G^\alpha(\omega)\left [ (2n_\alpha(\omega) + 1)\cos(\omega\tau) - i\sin(\omega\tau) \right ], \nonumber \\
\end{eqnarray}
where $G^\alpha(\omega)$ is the spectral density function, and $n_\alpha(\omega)$ is the average phonon occupation number for the $\alpha-$bath, characterized by the temperature $T_\alpha$.  $[\mu,\nu,\kappa,\lambda]$ label the eigenstates of $H_S$, and $\omega_{\mu,\nu}$ is the difference between eigenvalues $\epsilon_\mu$ and $\epsilon_\nu$. 
Specific functional forms for $S$ are described below.

The spectral density function of each bath is taken to be of the Ohmic form with an exponential cutoff,
\begin{eqnarray}
G^\alpha(\omega) = \gamma_\alpha \omega e^{-\omega/\omega_c}.
\end{eqnarray}
The golden rule rates in Eq. \ref{eqn:gamma_tensors} can be solved in closed form in the Markovian limit to give
\begin{eqnarray}
^\alpha\Gamma^{\pm}_{\lambda,\nu,\mu,\kappa} &=& \langle \lambda|S^\alpha |\nu \rangle \langle \mu|S^\alpha |\kappa \rangle \Upsilon^\alpha(\omega) + i\Delta^\alpha(\omega),
\label{eqn:RF_tensors_2}
\end{eqnarray}
where  $\Delta^\alpha$ is the Lamb shift, and 
\begin{eqnarray}
\Upsilon^{\alpha}(\omega) &=& \Bigg\{\begin{matrix}
\frac{1}{2}G^\alpha(|\omega|)n_\alpha(|\omega|;T_\alpha) & \omega < 0 \\ 
\frac{1}{2}G^\alpha(\omega)\left[ n_\alpha(\omega;T_\alpha) + 1\right ] & \omega > 0.
\end{matrix}
\end{eqnarray}\\

The heat transfer is characterized by the rate of heat exchange with the $hot$-bath,
\begin{eqnarray}
J_{H} = \textrm{Tr}\left [H_S{\cal D}^{H} \rhoss\right],
\label{eqn:J_q}
\end{eqnarray}
where ${\cal D}^{H}$ is the dissipator of the $hot$-bath, Eqs. \ref{eqn:Redfield}--\ref{eqn:Redfield_tensors}.

We also computed the rectification coefficient,
\begin{eqnarray}
R = \frac{|J_{H}| - |J'_{H}|}{|J_{H}| + |J'_{H}|},
\label{eqn:R}
\end{eqnarray}
where $J_{H}$ is the heat current when the hot bath interacts with site one and the cold bath with site two, while $J'_{H}$ is computed by swapping the temperatures of the H and C baths.
As we illustrate in the main draft, both functions, $J_{H}$ and $R$, can be maximized using gradient ascent based methods. 

\subsection{\label{sec:Lindblad}Energy transfer model}
The quantum master equation for a donor-acceptor exciton composite system is,
\begin{eqnarray}
\frac{\partial \hat{\rho}_{S}}{\partial t} = \cL_{0}[\hatrho] + \cL_{rad}[\hatrho] + \cL_{deph}[\hatrho] + \cL_{rec}[\hatrho] + \cL_{RC}[\hatrho],
\label{eqn:qme_KM}
\end{eqnarray}
where $\cL_{0}[\hatrho] = -i \left [ \hat{H}_{S}, \hatrho_{S}\right ]$ and $\hat{H}_{S}$ is the Hamiltonian of the system of interest. 
For this excitonic system, 
\begin{eqnarray}
H_S = \sum_{i=1}^2 \epsilon_i \ket{i}\bra{i} + J \big ( \ket{1}\bra{2} + \ket{2}\bra{1} \big ),
\end{eqnarray}
where $\epsilon_i$ are the site-energies for each state $\ket{i}$, and $J$ is the hopping constant \cite{Jung_JCP,TimurBrumer_JCP_148,Timur_PRL}. The energy eigenstates of $H_{S}$ are $\ket{e_{\pm}}$; $H_S \ket{e_{\pm}} = \epsilon_{\pm} \ket{e_{\pm}}$.

The remaining operators in Eq. \ref{eqn:qme_KM}, $\cL_{i}[\cdot]$, describe the radiation (rad), the trapping  of  the  excitons  at the  reaction  center (RC), environmental dephasing (deph), and the recombination of the excitons (rec). Each of the bath operators, except the radiative contribution, are taken to be of the Lindblad form
\begin{equation}
\label{eq:Ldephase}
\cL_i[\hatrho] = 2K_i\sum^2_{j=1} \left( L_i\hat{\rho}L_i^{\dagger}  -\frac{1}{2}[L_i^{\dagger}L_i, \hat{\rho}]_+\right),
\end{equation}
where $[A,B]_{+}$ is the anticommutator of $A$ and $B$, and $K_i$ are the dephasing rates. For $\cL_{deph}$ we take $K_{deph}=\gamma_d$ as the phonon bath dephasing rate and $L_{deph}=|k\rangle \langle k |$. For $\cL_{rec}$ the dephasing corresponding to the recombination rate $K_{rec}=\Gamma$ and $L_{rec}=|g\rangle \langle k |$. Lastly, $K_{RC}=\Gamma_{RC}$ and $L_{RC}=|RC\rangle \langle 2 |$.
The light-matter interaction term, $\cL_{rad}$, derived from the partially-secular Redfield approximation \cite{Timur_PRL,Timur_JCP,TimurBrumer_JCP_148} and is written in the eigenbasis of $H_{S}$ as
{\small
\begin{eqnarray}
\label{eq:Lrad}
(\cL_{\textrm{rad}}\rho)_{e_{\pm}e_{\pm}}  &=& -(r_{e_{\pm}} + \gamma_{e_{\pm}})\rho_{e_{\pm}e_{\pm}} + r_{e_{\pm}}\rho_{gg} \nonumber \\ 
& & - (\sqrt{r_{e_{+}}r_{e_{-}} } + \sqrt{\gamma_{e_{+}}\gamma_{e_{-}} })\rho^{R}_{e_{+}e_{-}} \\
(\cL_{\textrm{rad}}\rho)_{e_{+}e_{-}} &=& -\frac{1}{2}(r_{e_{+}} + \gamma_{e_{+}} +r_{e_{-}} + \gamma_{e_{-}} + 2i\Delta)\rho_{e_{+}e_{-}}  \nonumber \\
& & - \frac{1}{2}(\sqrt{r_{e_{+}}r_{e_{-}} } + \sqrt{\gamma_{e_{+}}\gamma_{e_{-}} })(\rho_{e_{+}e_{+}}+\rho_{e_{-}e_{-}}) \nonumber \\
& & + \sqrt{r_{e_{+}}r_{e_{-}} }\rho_{gg}, 
\end{eqnarray}}
where $\Delta = \sqrt{(\epsilon_1 - \epsilon_2)^2 - 4J^2}$ is the excitonic splitting,  $\gamma_{\pm}$ are the incoherent emission rates, and $r_\pm$ are the pumping rates for each eigenstate of $H_{S}$.
The total quantum master equation is,
{\small
\begin{eqnarray}
\label{eqn:ME_KennyModel}
\dot{\rho}_{e_{+}e_{+}} & = & -[2\Gamma + \gamma_d\sin^2(2\theta)+2\Gamma_{RC}\cos^2(\theta)]\rho_{e_{+}e_{+}} \nonumber  \\ 
& & +\gamma_d\sin^2(2\theta)\rho_{e_{-}e_{-}} + r_{e_{+}}\rho_{gg} \nonumber  \\ 
& & +[\Gamma_{RC}\sin(2\theta) - 2\gamma_d\sin(2\theta)\cos(2\theta)]\rho^{R}_{e_+e_-}, \nonumber \\
\dot{\rho}_{e_{-}e_{-}} & = & -[2\Gamma + \gamma_d\sin^2(2\theta)+2\Gamma_{RC}\sin^2(\theta)]\rho_{e_{-}e_{-}}\nonumber  \\ 
& & +\gamma_d\sin^2(2\theta)\rho_{e_{+}e_{+}} + r_{e_{-}}\rho_{gg} \nonumber  \\ 
& & +[\Gamma_{RC}\sin(2\theta) + 2\gamma_d\sin(2\theta)\cos(2\theta)]\rho^{R}_{e_+e_-}, \nonumber  \\
\dot{\rho}^{R}_{e_{+}e_{-}} & = & -[2\Gamma + \Gamma_{RC} + 2\gamma_d(1-\sin^2(2\theta))]\rho^{R}_{e_{+}e_{-}}\nonumber \\ 
& & +\Delta{\rho}^{I}_{e_{+}e_{-}} + \sqrt{r_{e_{+}} r_{e_{-}}}\rho_{gg} \nonumber \\
& & +\frac{1}{2}\Gamma_{RC}\sin(2\theta)(\rho_{e_{+}e_{+}} + \rho_{e_{-}e_{-}}) + \gamma_d\sin(2\theta)\cos(2\theta)(\rho_{e_-{e_-}}-\rho_{e_+{e_+}}),\nonumber \\
\dot{\rho}^{I}_{e_{+}e_{-}} & = & -[2\Gamma + 2\gamma_d + \Gamma_{RC}]\rho^{I}_{e_{+}e_{-}}  -\Delta\rho^{R}_{e_{+}e_{-}} ,
\end{eqnarray}
}
with $r_{+} = 2r\cos^2(\theta)$ and $r_{-} = 2r\sin^2(\theta)$,  and $\tan(2\theta) = 2J/(\epsilon_1-\epsilon_2)$. 
Additional details pertaining to this model can be found in Ref. \cite{Jung_JCP}. \\

In energy transfer scenarios one needs a metric to characterize the amount of energy transferred through the system. This is quantified by the exciton efficiency
\begin{eqnarray}
\eta_{loc} &=& \frac{\Gamma_{RC}}{r} \;\rho_{2}^{ss}.
\label{eqn:eta_loc}
\end{eqnarray}
Clearly $\eta_{loc}$ is a function of the parameters of the quantum master equation, Eq. \ref{eqn:ME_KennyModel}, and $\rho^{ss}_{2}$ is the density matrix of site $\ket{2}$ in the steady state limit. 

To optimize $\eta_{loc}$ with gradient descent we require the knowledge of $\frac{\partial \rho_{2}^{ss}}{\partial \Theta_i}$, where $\Theta_i$ is any parameter of the quantum master equation. \\

A code for the energy transfer system is publicly available through the following repository: \url{https://github.com/RodrigoAVargasHdz/steady_state_jax}.

\section{\label{sec:Results}Results}
In this section we present the values of the optimal models for a quantum heat transfer model. 

The most general type of interactions $S^\alpha$ (Eq.\ref{eqn:RF_tensors_2}) with the hot (H), cold (C) and decoherence (D) baths are,
\begin{eqnarray}
S^H &=& a_0\ket{g}\bra{1} + a_1 \ket{g}\bra{2} + \textrm{h.c.} \label{eqn:_S_H}\\
S^C &=& b_0\ket{g}\bra{1} + b_1 \ket{g}\bra{2} + \textrm{h.c.}\label{eqn:_S_C}\\
S^D &=& \ket{1}\bra{2} + \textrm{ h.c.} \label{eqn:_S_D},
\end{eqnarray}
where $[a_0,a_1]$ and $[b_0,b_1]$ are the coupling strength parameters to the H and C bath. 
The $D-$bath is a control mechanism to study the role of coherences between the sites.
Analytic results for $J_\alpha$ can only be derived in the secular limit \cite{Dvira_PRE}. 

For the simulations we fixed $T_{H} = 0.15$, $T_{C} = 0.1$, and $T_{D} = 0.12$. 
We use the Adam optimizer with a learning rate = 0.01 to optimize $J_{H}$.
To ensure the values of $a_0$, $a_1$, $b_0$, and $b_1$ are between [0,1], we used the softmax function. $\theta$ and $J$ were restricted to only positive values by an exponential function.
We report the collection of optimal parameters in Table \ref{tab:QHE_J_H}.
For Table \ref{tab:QHE_J_H}, we can find that the optimal models could be categorize in two groups, where the hot and cold bath mainly interact with independent sites, (i) $a_{0}=b_{1}\approx 1$ and $a_{1}=b_{0}\approx 0$ or vice versa, and (ii) when both hot and cold bath mainly interact with one site   $a_{i}=b_{i}\approx 1$ and $a_{j}=b_{j}\approx 0$. This trend is discussed in greater detail in the main text.

\begin{table}[h!]
\caption{\label{tab:QHE_J_H} Optimal parameters for a quantum heat transfer model found by maximizing $J_{H}$ with Adam. For each set of parameters, the initial values were sampled randomly. We held fixed  $T_{H} = 0.15$, $T_{C} = 0.1$, and $T_{D} = 0.12$.  }
    \vspace{0.25cm}
    \centering
\begin{tabular}{ c|c| c c c |c c | c c | c } 
 $\theta$ & $J$ & $\gamma_{H}$ & $\gamma_{C}$ & $\gamma_{D}$ & $a_0$ & $a_1$ & $b_0$ & $b_1$ & $J_{H}$   \\  \hline \hline
 0.3900 &  0.0164 & 0.0021 & 0.0020 &  0.0001 & 0.9995 & 0.0005 & 0.0006 & 0.9994 & 1.598e-05 \\
 0.3890 &  0.0172 & 0.0022 & 0.0024 &  0.0001 & 0.0001 & 0.9999 & 0.9998 & 0.0002 & 1.789e-05 \\
 0.3897 &  0.0044 & 0.0025 & 0.0023 &  0.0001 & 0.0007 & 0.9993 & 0.0001 & 0.9999 & 1.867e-05 \\
 0.4022 &  0.0664 & 0.0025 & 0.0025 &  0.0004 & 0.9999 & 0.0001 & 0.9999 & 0.0001 & 1.831e-05 \\
 0.3891 &  0.0180 & 0.0025 & 0.0025 &  0.0002 & 0.0021 & 0.9979 & 0.9999 & 0.0001 & 1.920e-05 \\
 0.3896 &  0.0041 & 0.0025 & 0.0024 &  0.0018 & 0.0002 & 0.9998 & 0.0001 & 0.9999 & 1.882e-05 \\
 0.3898 &  0.0177 & 0.0025 & 0.0024 &  0.0002 & 0.9999 & 0.0001 & 0.0002 & 0.9998 & 1.869e-05 \\
 0.3888 &  0.0052 & 0.0024 & 0.0025 &  0.0001 & 0.9992 & 0.0008 & 0.9998 & 0.0002 & 1.908e-05 \\
 0.3895 &  0.0126 & 0.0025 & 0.0025 &  0.0001 & 0.9992 & 0.0008 & 0.9999 & 0.0001 & 1.934e-05 \\
 0.3893 &  0.0178 & 0.0025 & 0.0025 &  0.0001 & 0.0002 & 0.9998 & 0.9998 & 0.0002 & 1.927e-05 \\
 0.3892 &  0.0124 & 0.0024 & 0.0025 &  0.0014 & 0.9998 & 0.0002 & 0.9994 & 0.0006 & 1.910e-05 \\
 0.3897 &  0.0041 & 0.0025 & 0.0024 &  0.0019 & 0.9998 & 0.0002 & 0.9999 & 0.0001 & 1.888e-05 \\
 0.3895 &  0.0178 & 0.0025 & 0.0024 &  0.0001 & 0.9994 & 0.0006 & 0.0002 & 0.9998 & 1.904e-05 \\
 0.3885 &  0.0174 & 0.0022 & 0.0025 &  0.0001 & 0.0003 & 0.9997 & 0.9989 & 0.0011 & 1.807e-05 \\
 0.3894 &  0.0060 & 0.0024 & 0.0023 &  0.0004 & 0.9998 & 0.0002 & 0.9998 & 0.0002 & 1.821e-05 \\
 0.3899 &  0.0190 & 0.0025 & 0.0024 &  0.0008 & 0.9998 & 0.0002 & 0.0004 & 0.9996 & 1.884e-05 \\
 0.3893 &  0.0085 & 0.0025 & 0.0025 &  0.0021 & 0.9990 & 0.0010 & 0.9999 & 0.0001 & 1.938e-05 \\
 0.3904 &  0.0178 & 0.0025 & 0.0022 &  0.0005 & 0.9998 & 0.0002 & 0.0002 & 0.9998 & 1.816e-05 \\
 0.4032 &  0.0651 & 0.0025 & 0.0024 &  0.0021 & 0.0015 & 0.9985 & 0.9998 & 0.0002 & 1.811e-05 \\
 \hline
\end{tabular}
\end{table}

For the quantum heat transfer system were we maximize the rectification coefficient, we optimized $\theta, J,\gamma_{H}, \gamma_{C}$, and $\gamma_{D}$. 
We held  fixed $T_{H} = 0.15$, $T_{C} = 0.1$, and $T_{D} = 0.12$.
We use the Adam optimizer with a learning rate = 0.01. For these simulations the hot and cold bath only interact with opposite sites. 
$\theta$ and $J$ were restricted to only positive values by an exponential function.
We report the collection of optimal parameters in Table \ref{tab:QHE_R}.

From Table \ref{tab:QHE_R}, we can observe that for values where $R > 0.95$, there exists a strong linear correlation between the energy of sites $\ket{1}$ and $\ket{2}$ $\theta$, and the hopping parameter $J$; Figure 3 in the main text.

\begin{table}[h!]
    \centering
\caption{\label{tab:QHE_R} Optimal parameters for a quantum heat transfer model found by maximizing the rectification coefficient, Eq. \ref{eqn:R}, with Adam. For each set of parameters, the initial values were sampled randomly and the optimization was stopped once $R > 0.95$. For these calculations, we held fixed $T_{H} = 0.15$, $T_{C} = 0.1$, and $T_{D} = 0.12$. } 
\begin{tabular}{ c|c| c c c}
 $\theta$ & $J$ & $\gamma_{H}$ & $\gamma_{C}$ & $\gamma_{D}$    \\  \hline \hline
      0.190 &      0.111 &     0.0002 &     0.0012 &     0.0004 \\
     0.313 &      0.052 &     0.0001 &     0.0007 &     0.0000 \\
     0.502 &      0.102 &     0.0017 &     0.0002 &     0.0001 \\
     0.875 &      0.594 &     0.0002 &     0.0008 &     0.0009 \\
     0.127 &      0.043 &     0.0008 &     0.0017 &     0.0007 \\
     0.001 &      0.150 &     0.0004 &     0.0025 &     0.0023 \\
     0.516 &      0.089 &     0.0009 &     0.0002 &     0.0020 \\
     0.146 &      0.083 &     0.0002 &     0.0015 &     0.0010 \\
     0.265 &      0.087 &     0.0013 &     0.0007 &     0.0008 \\
     0.233 &      0.108 &     0.0003 &     0.0010 &     0.0002 \\
     0.685 &      0.441 &     0.0007 &     0.0008 &     0.0002 \\
     0.281 &      0.356 &     0.0008 &     0.0021 &     0.0003 \\
     0.993 &      0.330 &     0.0001 &     0.0001 &     0.0015 \\
     0.997 &      0.781 &     0.0002 &     0.0013 &     0.0001 \\
     0.880 &      0.669 &     0.0000 &     0.0013 &     0.0020 \\
     0.359 &      0.083 &     0.0004 &     0.0005 &     0.0021 \\
     0.128 &      0.071 &     0.0000 &     0.0017 &     0.0007 \\
     0.506 &      0.218 &     0.0012 &     0.0004 &     0.0012 \\
     0.382 &      0.116 &     0.0000 &     0.0005 &     0.0001 \\
     0.635 &      0.175 &     0.0005 &     0.0002 &     0.0008 \\
     0.656 &      0.210 &     0.0018 &     0.0001 &     0.0022 \\
     0.082 &      0.034 &     0.0009 &     0.0022 &     0.0020 \\
     0.797 &      0.259 &     0.0015 &     0.0001 &     0.0013 \\
     0.902 &      0.494 &     0.0007 &     0.0003 &     0.0023 \\
  \hline
\end{tabular}
\end{table}

\break
\subsection{Energy transfer model}
The optimal collection of parameters depicted in Fig. 4 in the main manuscript are reported in Table \ref{tab:KennyModel}.

We maximize the exciton steady state efficiency ($\eta_{loc}$) by optimizing $\vTheta=[\Gamma,\gamma_d],J,|\epsilon_1 - \epsilon_2|]$, where  $\Gamma$ is the recombination rate, $\gamma_d$ is the phonon bath dephasing rate, $J$ is the hopping amplitude, and  $|\epsilon_1 - \epsilon_2|$ is the energy difference between site $\ket{1}$ and $\ket{2}$.
From Table \ref{tab:KennyModel}, we can observe that the optimal values of $\Gamma$ span in the range between $10^{-3.8}$ and  $10^{-3.17}$ ps$^{-1}$. We can also observe that $J$ and $|\epsilon_1 - \epsilon_2|$  are linearly correlated when $|\epsilon_1 - \epsilon_2| < 0.5$ ps$^{-1}$ and $J < 3.0$ ps$^{-1}$.
Furthermore, $\gamma_d$ is the parameter that span the largest range for efficient steady state transport, $10^{9} < \gamma_d < 10^{13}$ Hz.
Each independent optimization was stopped when the steady state efficiency was greater than 99$\%$, which for the majority of the trajectories took approximately 20 iterations with Adam.

\begin{table}[h!]
    \caption{\label{tab:KennyModel} Optimal parameters for a quantum energy transfer model found by maximizing the exciton steady state efficiency ($\eta_{loc}$), Eq. \ref{eqn:eta_loc}, with Adam.  
    For each set of parameters, the initial values were sampled randomly. We held fixed  $\Gamma_{RC} = 0.1$ ps${-1}$, and $r = 6.34\;\times\;10^{-10}$  ps${-1}$ \cite{Timur_JCP}.  Each individual optimization was stopped when $\eta_{loc} > 0.99$.} \vspace{0.25cm}
\begin{minipage}{0.40\textwidth}
    {\scriptsize
    \begin{tabular}{c| c |c c}
    J(ps$^{-1}$) & $|\epsilon_{1}-\epsilon_{2}|$ (ps$^{-1}$) &  $log(\gamma_{d})$ (Hz) & $log(\Gamma)$ (ps$^{-1}$) \\ \hline\hline
     2.690 &      0.257 &     26.353 &     -7.613 \\ 
     2.399 &      0.194 &     25.021 &     -8.206 \\ 
     2.652 &      0.326 &     25.593 &     -7.974 \\ 
     2.659 &      0.266 &     25.682 &     -7.801 \\ 
     3.184 &      0.889 &     27.915 &     -7.330 \\ 
     2.642 &      0.327 &     26.054 &     -7.998 \\ 
     2.317 &      0.118 &     20.826 &     -8.536 \\ 
     2.658 &      0.266 &     25.606 &     -7.818 \\ 
     2.373 &      0.175 &     24.383 &     -8.268 \\ 
     2.649 &      0.312 &     25.176 &     -7.954 \\ 
     2.645 &      0.267 &     25.134 &     -7.893 \\ 
     2.825 &      0.238 &     27.129 &     -7.374 \\ 
     2.340 &      0.145 &     22.894 &     -8.396 \\ 
     4.100 &      3.304 &     29.026 &     -7.713 \\ 
     4.511 &      1.576 &     29.752 &     -8.094 \\ 
     3.245 &      0.760 &     27.998 &     -7.345 \\ 
     2.463 &      0.239 &     26.003 &     -8.106 \\ 
     2.406 &      0.200 &     25.192 &     -8.190 \\ 
     2.682 &      0.260 &     26.234 &     -7.650 \\ 
     2.655 &      0.267 &     25.494 &     -7.840 \\ 
     2.698 &      0.253 &     26.430 &     -7.588 \\ 
     2.713 &      0.255 &     26.567 &     -7.542 \\ 
     2.451 &      0.232 &     25.886 &     -8.119 \\ 
     2.828 &      0.235 &     27.139 &     -7.372 \\ 
     3.217 &      0.413 &     27.955 &     -7.335 \\  \hline
    \end{tabular}}
\end{minipage} \hfill
\begin{minipage}{0.40\textwidth}
    {\scriptsize
    \begin{tabular}{c| c |c c}
    J(ps$^{-1}$) & $|\epsilon_{1}-\epsilon_{2}|$ (ps$^{-1}$) &  $log(\gamma_{d})$ (Hz) & $log(\Gamma)$ (ps$^{-1}$) \\ \hline\hline
     2.582 &      0.303 &     26.451 &     -8.035 \\ 
     2.460 &      0.238 &     25.983 &     -8.108 \\ 
     3.246 &      3.596 &     29.266 &     -8.549 \\ 
     2.554 &      0.289 &     26.449 &     -8.047 \\ 
     2.631 &      0.271 &     24.032 &     -7.928 \\ 
     2.785 &      0.255 &     26.979 &     -7.414 \\ 
     2.653 &      0.323 &     25.470 &     -7.968 \\ 
     4.203 &      3.193 &     29.265 &     -7.867 \\ 
     2.541 &      0.284 &     26.433 &     -8.053 \\ 
     2.471 &      0.245 &     26.084 &     -8.097 \\ 
     3.295 &      2.166 &     28.055 &     -7.364 \\ 
     2.645 &      0.267 &     25.150 &     -7.892 \\ 
     2.328 &      0.132 &     22.032 &     -8.461 \\ 
     2.315 &      0.117 &     20.767 &     -8.539 \\ 
     2.923 &      0.324 &     27.427 &     -7.321 \\ 
     2.572 &      0.299 &     26.459 &     -8.039 \\ 
     2.564 &      0.295 &     26.457 &     -8.043 \\ 
     3.441 &      4.014 &     29.169 &     -8.419 \\ 
     2.996 &      4.876 &     29.022 &     -8.729 \\ 
     4.039 &      1.448 &     28.974 &     -7.691 \\ 
     2.335 &      0.140 &     22.619 &     -8.418 \\ 
     2.608 &      0.314 &     26.371 &     -8.022 \\ 
     2.397 &      0.192 &     24.946 &     -8.213 \\ 
     2.348 &      0.151 &     23.250 &     -8.368 \\ 
     2.430 &      0.217 &     25.610 &     -8.147 \\ \hline
\end{tabular}}
\end{minipage}
\end{table}

\bibliography{references}